\documentclass[
    ,final            
  ]
  {aipproc}

\layoutstyle{6x9}

\newcommand{\lsim}{\raise.3ex\hbox{$<$\kern-.75em\lower1ex\hbox{$\sim$}}}

\newcommand{\be}{\begin{equation}}
\newcommand{\ee}{\end{equation}}

\begin{document}
\hspace{9.8 cm}FZJ--IKP(TH)--2003--05

\title {Investigating scalar meson mixing in nucleon--nucleon and
  deuteron--deuteron collisions\footnote{Invited talk at the international
    High Energy Physics Workshop {\it Scalar Mesons: an Interesting Puzzle for
      QCD}, May 16-18, 2003, SUNY Institute of Technology, Utica, New York.}}

\classification{43.35.Ei, 78.60.Mq}
\keywords{Document processing, Class file writing, \LaTeXe{}}

\author{C. Hanhart}{
  address={IKP, Forschungszentrum J\"ulich, 52428 J\"ulich, Germany},
  email={c.hanhart@fz-juelich.de},
}

%

\copyrightyear  {2001}

\begin{abstract}
The advantages to study charge--symmetry breaking and especially
the phenomenon of $f_0-a_0$ mixing in nucleon--nucleon and
deuteron--deuteron induced reactions are discussed.
\end{abstract}

\date{\today}

\maketitle

\section{Introduction}

As is clearly demonstrated by the existence of these proceedings,
the scalar mesons still are an interesting puzzle with respect to
our understanding of QCD. 
It is striking that, although experimentally well established since
many years and seen in various reactions, the fundamental structure
of the $f_0 (980)$ and the $a_0(980)$ is still obscure.

Already long ago Achasov argued \cite{achasov}
\begin{itemize}
\item that the almost degenerate isoscalar meson $f_0(980)$
 and the isovector meson $a_0(980)$ should mix and
\item that, in the vicinity of 
the $K\bar K$ threshold, this mixing should be enhanced by an order of magnitude compared to
usual charge--symmetry breaking (CSB) effects.
\end{itemize}
However, up to now no unambiguous experimental signal of this predicted
behaviour has been seen.
We will argue that the ideal reactions to extract
information of this CSB matrix element is the production of $f_0$ and $a_0$ in
$NN$ and $dd$ collisions, since these reactions allow to manipulate the
isospin of the initial and final states such that CSB leads to unique
signals that can be extracted model independently. For details about the
planned experimental program we refer to Ref. \cite{buscher}.

The paper is organized as follows: in the next section we review what is
known about $a_0$--$f_0$ mixing. In section three the
particular features of the reactions $NN\to dX$ and $dd\to \alpha X$ are
discussed in detail. Section four contains a presentation of recent
experimental results for the reaction $pp\to d\bar K^0 K^+$ that show, that
the scalar resonances can indeed be studied in $NN$ collisions close to the
kaon threshold. The paper
closes with a brief summary as well as an outlook.

\section{Remarks about $f_0-a_0$ mixing}

\begin{figure}[t!]
  {\includegraphics[height=.12\textheight]{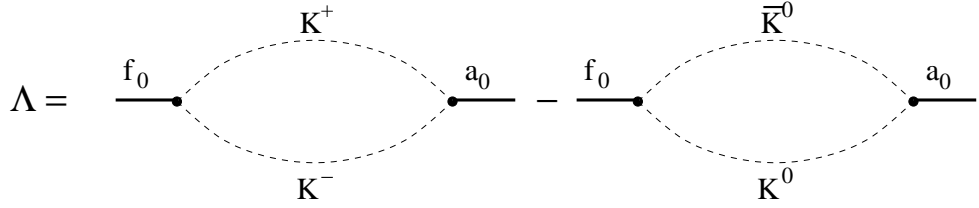}}
  \caption{Graphical illustration of the leading contribution to the $f_0-a_0$ mixing
    matrix element $\Lambda$ defined
    in Eq. (\protect\ref{llam}).}
\label{lcsb}
\end{figure}

\begin{figure}[t!]
  {\includegraphics[height=.4\textheight]{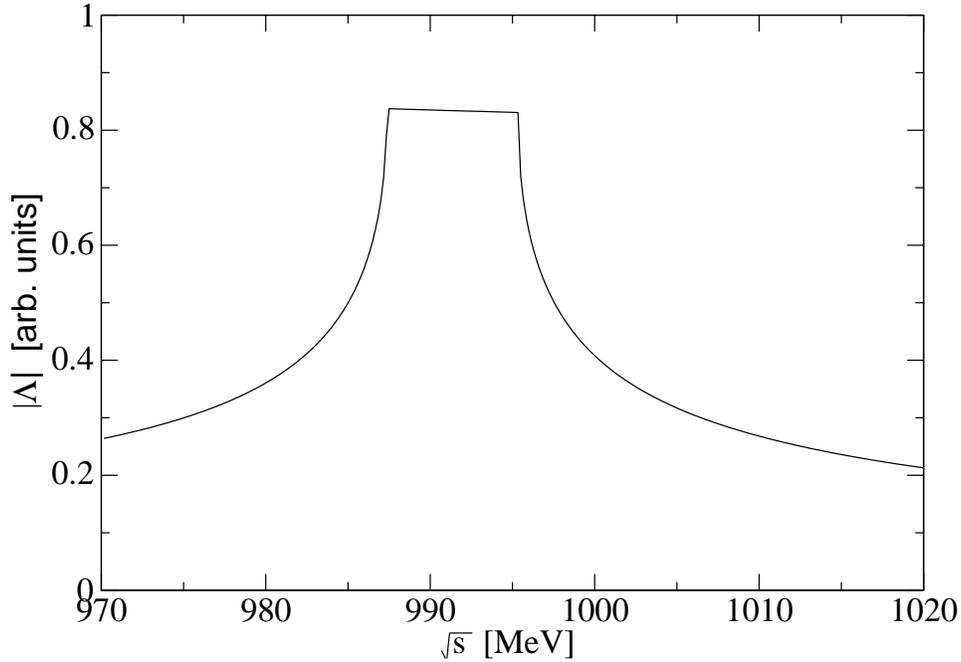}}
  \caption{Modulus of the leading piece of the mixing amplitude
    $\Lambda$ defined in Eq. (\protect\ref{llam}). The two kinks occur at the
     $K^+ K^-$ (at 987.35 MeV) and the $\bar K^0 K^0$ (995.34 MeV) threshold respectively.}
\label{mixsdep}
\end{figure}

In Ref. \cite{achasov} it was demonstrated, that the leading piece of the  $f_0-a_0$ mixing
amplitude
can be written as\footnote{Here we deviate from the original notation of
  Achasov et al. in order to introduce dimensionless coupling constants in
  line with the standard Flatt\'e parameterization.}
\begin{equation}
\Lambda = \langle f_0 |T| a_0\rangle = ig_{f_0K\bar K}g_{a_0K\bar
    K}\sqrt{s}\left( p_{K^0}-p_{K^+} \right) \ + {\cal
    O}\left(\frac{p_{K^0}^2-p_{K^+}^2}{s}\right) \ ,
\label{llam}
\end{equation}
where $p_K$ denotes the modulus of the relative momentum of the kaon pair and
the effective coupling constants are defined through $\Gamma_{xK\bar
  K}=g_{xK\bar K}^2p_K$. Obviously, this leading contribution is just that of
the unitarity cut of the diagrams shown in Fig. \ref{lcsb} and is therefore
model independent. In addition, the contribution shown in Eq.  (\ref{llam}) is
unusually enhanced between the $K^+ K^-$ and the $\bar K^0 K^0$ thresholds, a
regime of only 8 MeV width. Here it scales as
$$\sqrt{\frac{m_{K^+}^2-m_{K^0}^2}{m_{K^+}^2+m_{K^0}^2}}\sim
\sqrt{\frac{m_u-m_d}{m_u+m_d}}\ ,$$
where $m_u$ and $m_d$ denote the current quark mass of the up and down quark
respectively. This is
in contrast to common CSB
effects\footnote{Here we denote as common CSB effects those that occur at the
  Lagrangian level.} which scale as $(m_u-m_d)/(m_u+m_d)$, since they have to
be analytic in the quark masses. It is easy to see that away from the kaon
thresholds $\Lambda$ returns to a value of natural size.  This $\sqrt{s}$
dependence of $\Lambda$ is depicted in Fig. \ref{mixsdep}. Note, in Eq.
(\ref{llam}) electro magnetic effects were neglected for they are also
subleading \cite{achasov}.

So far, little is known about the effective coupling of the $f_0$ to kaons.
Values in the literature vary from 2.51 \cite{phidecay} down to values
compatible with zero \cite{ddecays}. An accurate measurement of $\Lambda$
therefore will  strongly constrain $g_{f_0K\bar K}$. It should be
stressed that in the couplings of physical particles to mesons important
information about the nature of that particle is contained, as was shown by
Weinberg for the case of the deuteron \cite{weinberg}.

To our knowledge the impact of the CSB
$f_0-a_0$ mixing on hadronic reactions close to the kaon threshold
was studied in one microscopic model only \cite{oli}. There, the 
CSB cross section $\pi^+\pi^- \to \pi^0\eta$ was
predicted
to be of the order of a few hundred $\mu b$ in line with the order of
magnitude
estimate derived from Eq. (\ref{llam})\footnote{Here we assume the effective
couplings to be of order 1.}.

It is important to stress that $f_0-a_0$ mixing might have a significant
impact also on the decay $\eta\to 3\pi$ \cite{schechter}, however, in a very
different kinematic regime compared to what we are looking at here. Thus one
should expect that in addition to kaon loops also other CSB effects play a
role. For a complete understanding of the $f_0-a_0$ mixing mechanisms 
knowledge about both kinematic regimes, that close to the resonance poles as
well as that  of the eta
decay, is necessary.

With  recent papers by Close and Kirk \cite{close1,close2} the interest in
$a_0$--$f_0$ mixing was revived. Based on an analysis of central $pp$
collisions
as well as radiative phi decays the authors extracted a mixing matrix element 
that was a factor of five larger than that given in Eq. (\ref{llam})  and
independent of the invariant mass of the system produced. If confirmed, such a
large mixing would indicate a completely different mechanism at work 
compared to what we believe in at present. Please note, that the
work of Close and Kirk was heavily criticized in the literature
\cite{achasovcrit,ollercrit}. In any case, further experimental information
is urgently called for.

In Ref. \cite{tabakin} it was stressed that a large mixing as that given by
Eq.  (\ref{llam}) should have a measurable impact on observables in
photoproduction of scalar mesons. However, the signal will show up only as a
modification of some observables and a good theoretical understanding of the
rest of the amplitude is required to extract the strength of the mixing matrix
element. In the next section we will argue that nucleon--nucleon induced
reactions are especially suited to get insights on CSB and especially the
$f_0-a_0$ mixing matrix element.

\section{The special features of $NN$ and $dd$ induced reactions}
\label{spfe}

For an unambiguous result for the mixing amplitude it is desirable to
study observables that vanish in the absence of CSB. To identify such
observables the production of scalars in $NN$ and $dd$ collisions is
especially suited since these systems allows to manipulate the isospin.
Especially when using deuterons as isospin filters, selection rules following
from the symmetries of the strong interaction (parity, isospin) strongly
restrict the behavior of various observables for these reactions. Therefore
it is straight forward to study the violation of such symmetries in $NN$ and $dd$
collisions \cite{csbrep}.

Experimentally so far the signals of CSB where extracted from $pn\to d \pi^0$
\cite{allenapi0} and $dd\to \alpha \pi^0$ \cite{dd2alphapi0}. Since both the
$\alpha$ particle as well as the deuteron are isoscalars whereas the pion is
an isovector it is obvious that the latter reaction can only happen in the
presence of CSB. To identify the signal from CSB in the former case one has to
acknowledge that a $pn$ state, when being in a definite isospin state as it
is in $pn\to d \pi^0$ (the final state projects on isospin 1) behaves
as a pair of identical particles as long as isospin is conserved. One obvious
consequence is that the differential pion production cross section needs to be
forward--backward symmetric. Thus any deviation from forward--backward
symmetry is a signal of CSB. Analogously, one can look at a
forward--backward asymmetry in $pn\to d a_0^0$ and the total cross section in
$dd\to \alpha a_0^0$ by measuring a $\pi \eta$ pair in the final state at
invariant masses of about 1 GeV \cite{sascha,veramix} to study the scalar
meson mixing. Note that $\pi \eta$ is the dominant decay channel of the $a_0$.

To extract and identify the relevant CSB mechanisms in the pion production
reactions a detailed analysis within effective field theory is needed. The
potentially most relevant sources of CSB currently discussed are isospin
violating pion--nucleon scattering through the so called Weinberg term and
$\pi-\eta$ mixing \cite{dd2alphapi0}.  However, for a quantitative
understanding of the cross sections measured still a lot of work from the
theory side is needed.

\begin{figure}[t!]
  {\includegraphics[height=.3\textheight]{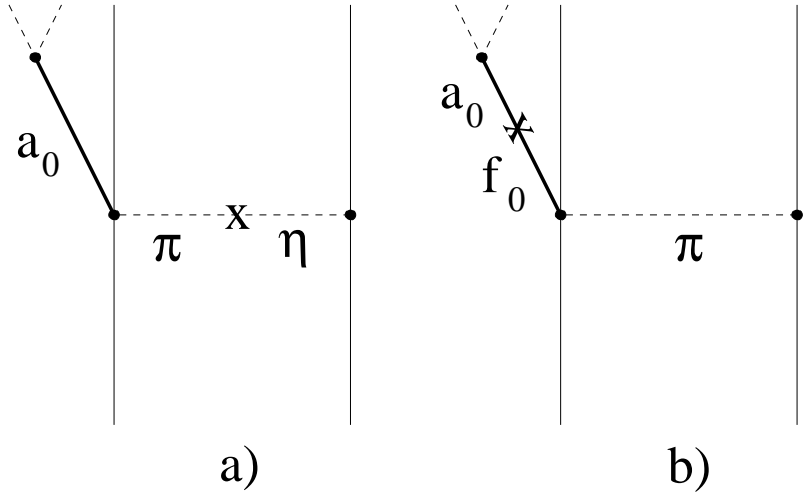}}
  \caption{Illustration of different sources of charge symmetry breaking:
    diagram a) shows CSB in the production operator through $\pi-\eta$ mixing
    and diagram b) shows CSB in the propagation of the scalars. Thin solid
    lines denote nucleons, thick solid lines scalar and dashed ones
    pseudoscalar mesons. The $X$ indicates the occurrence of a CSB
    matrix element.}
\label{diffcsb}
\end{figure}

In case of the production of scalar mesons the interpretation of a measured
signal of CSB is more straight forward, since the resonances $a_0$ and $f_0$
are overlapping (the values for masses and widths given in the current version
of the particle data booklet are basically equal \cite{pdb}). Thus it should
be intuitively clear that CSB in the propagation of the scalar mesons is
kinematically favored as compared to CSB in the production operator. To make
this statement more quantitative we compare the impact of $f_0-a_0$ mixing in
the propagation of the scalar mesons (Fig. \ref{diffcsb}b) to that of
$\pi-\eta$ mixing in the production operator (Fig. \ref{diffcsb}a). We regard
the latter as a typical CSB effect and thus as a reasonable order of magnitude
estimate for CSB in the production operator.  Note, that the relevant
dimensionless quantity for this comparison is the mixing matrix element times
a propagator (c.f. Fig. \ref{diffcsb}). In the production operator the
momentum transfer---at least close to the production threshold---is given by
$t=-M_Nm_R$, where $m_R$ denotes the invariant mass of the meson system
produced (or equivalently the mass of the resonance) and $M_N$ denotes the
nucleon mass. Thus, the appearance of the $\eta$ propagator introduces
a factor of about $1/t$ into the amplitude, since $t\gg m_\eta^2$.  On the other
hand,  the resonance propagator is given by $1/(m_R\Gamma_R)$,
as long as we concentrate on invariant masses of the outgoing meson system
close to the resonance position. Here
$\Gamma_R$ denotes the width of the scalar resonance. Thus we find using
$\Gamma_R= 50$ MeV, that the CSB in the production operator is kinematically
suppressed by a factor of more than $\Gamma_R/M_N \sim 1/20$ as compared to CSB
in the propagation of the scalars. In addition, as was argued in the first
section, the mixing matrix element of $f_0-a_0$ mixing as it occurs in the
propagation of the scalars is enhanced.  Given these two arguments we can be
sure that CSB as it will be measured in $NN$ and $dd$ collisions close to the
$K\bar K$ threshold will be dominated by $f_0-a_0$ mixing in the propagation
of the scalars and we can assume the production operator as charge symmetry
conserving.

There are two more reasons why looking at  $f_0-a_0$ mixing in
 $pn\to d\pi^0\eta$ is
particularly suited:
\begin{itemize}
\item the signal from mixing is kinematically enhanced \cite{sascha} and
\item the analyzing power gives a striking signal as well \cite{unsers1,unsers2}.
\end{itemize}
To understand the first point we have to look in a little more detail into the
selection rules for the reaction under discussion. If we assume isospin
conservation the $pn$ system in the initial state has to fulfill the
Pauli--Principle requiring $L+S+T$ being odd, where $L$, $S$ and $T$ denote
the total angular momentum, spin and isospin of the initial state. Isospin
conservation demands for an isovector (isoscalar) final state $L+S$ to be of even
(odd) in the initial state.  If we assume all final systems (the meson--meson
system produced as well as the deuteron with respect to this) in an $s$--wave,
then the final state has even parity and total angular momentum $J=1$, since
the deuteron has $J_d=1$.  Then parity conservation demands the initial state
to be even parity as well and thus we find that the overall $s$--wave in the
final state calls for even parity in the initial state. Thus the production of
an isovector needs an even parity $S=0$ state with $J=0$ as initial
state---this does not exist.  However, in the presence of CSB the $\pi \eta$
system might stem from an $f_0$ that in the propagation converts into an $a_0$.
In this case the selection rules for the production of an isoscalar apply,
calling for a $S=1$ even parity state with $J=1$. Thus both $^3S_1$ and
$^3D_1$ are possible initial states leading to $s$--wave production of a
scalar together with the deuteron. Therefore for kinematics close to the
resonance poles the charge symmetry allowed amplitude is suppressed by a
centrifugal barrier, whereas the charge--symmetry forbidden one is not.

To see that polarization observables and especially the analyzing power might
be used as well one has to study in detail the corresponding amplitude
structure. For this discussion we refer to Ref. \cite{unsers1}. However, it
should be stressed that in contrast to the forward--backward asymmetry of the
$\pi\eta$ system with respect to the deuteron, that can only occur in the
presence of CSB, the corresponding signal in the analyzing power can also come
from $p$--waves in the meson--meson system \cite{unsers2}.  Fortunately one
can study the contamination from meson--meson $p$--waves experimentally by
studying the analyzing power in the corresponding charged channels ($\vec
pp\to d\bar K^0 K^+$ and $\vec pp\to d \pi^+ \eta$) where there is no CSB. In
this context it is reassuring that the $K\bar K$ $p$--waves in the reaction
$pp\to d\bar K^0 K^+$ are found to be small (c.f.  next section) and the $\eta
\pi$ $p$--waves are estimated to be small \cite{unsers2,oset}.

\section{The reaction $pp\to d\bar K^0 K^+$}

\begin{figure}[t!]
  {\includegraphics[height=.3\textheight]{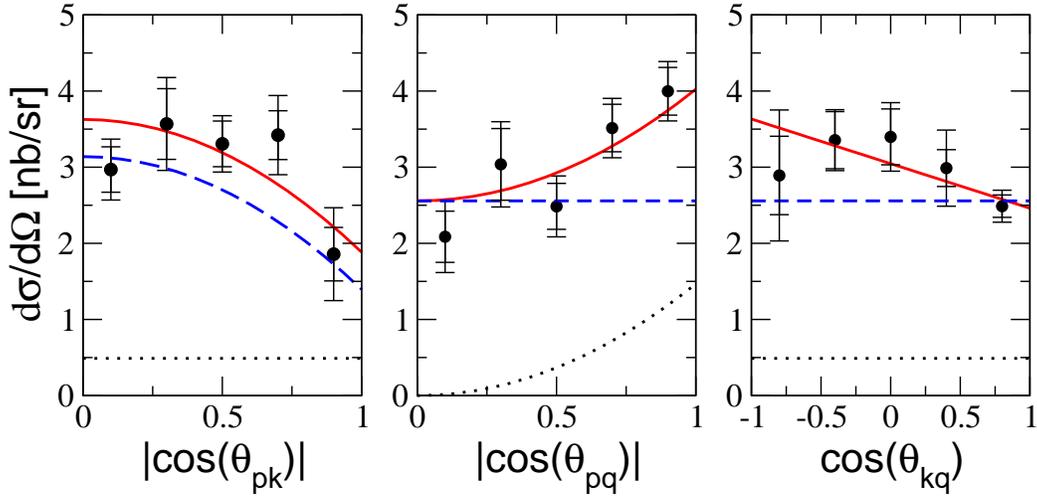}}
  \caption{Angular distributions for the reaction
    $pp\to d\bar K^0 K^+$.  The solid line shows the result of the overall
    fit including both $\bar K K$ $s$--wave as well as $p$--wave. To obtain
    the dashed (dotted) line the parameters for the $p$--wave ($s$--wave)
    where set to zero (see text).   The small error bars
show the statistical uncertainty only, whereas the large ones contain both
    the systematic as well as the statistical uncertainty (c.f. Ref.
    \protect\cite{a0exp}).}
\label{andist}
\end{figure}

Recently a first measurement of the reaction $pp\to d\bar K^0 K^+$ was
reported close to the two kaon threshold \cite{a0exp} (the measurement was
performed at an excess energy of only 46 MeV). In this section we will argue
that, based on the experimental evidence together with very general
assumptions, one has to conclude that the reaction $pp\to d\bar K^0 K^+$ is
dominated by the scalar meson production in the final state. Thus, it is
possible to study the scalar mesons in $NN$ induced reactions. This
observation is not at all trivial, for there is another potentially strong
final state interaction, namely that of the $\bar K^0$ with the deuteron that
is enhanced due to the existence of the $\Lambda (1405)$, as was stressed in
Ref. \cite{oset}.

\begin{figure}[t!]
  {\includegraphics[height=.4\textheight]{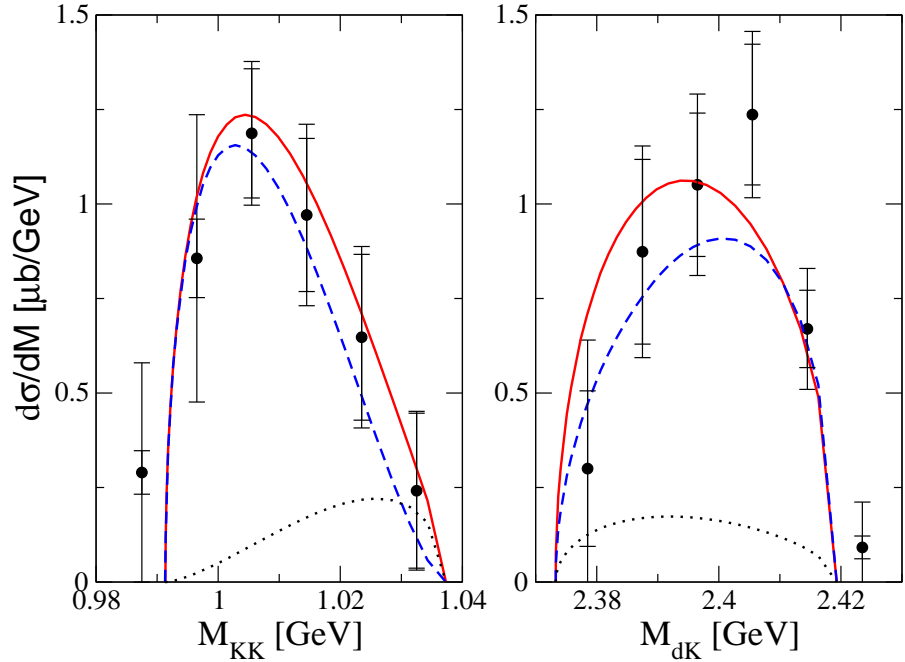}}
  \caption{Various mass distributions for the reaction $pp\to d\bar K^0 K^+$.
    Line code as in Fig. \protect\ref{andist}.  The small error bars
show the statistical uncertainty only, whereas the large ones contain both
    the systematic as well as the statistical uncertainty (c.f. Ref.
    \protect\cite{a0exp}).}
\label{mdist}
\end{figure}

\begin{figure}[t!]
  {\includegraphics[height=.4\textheight]{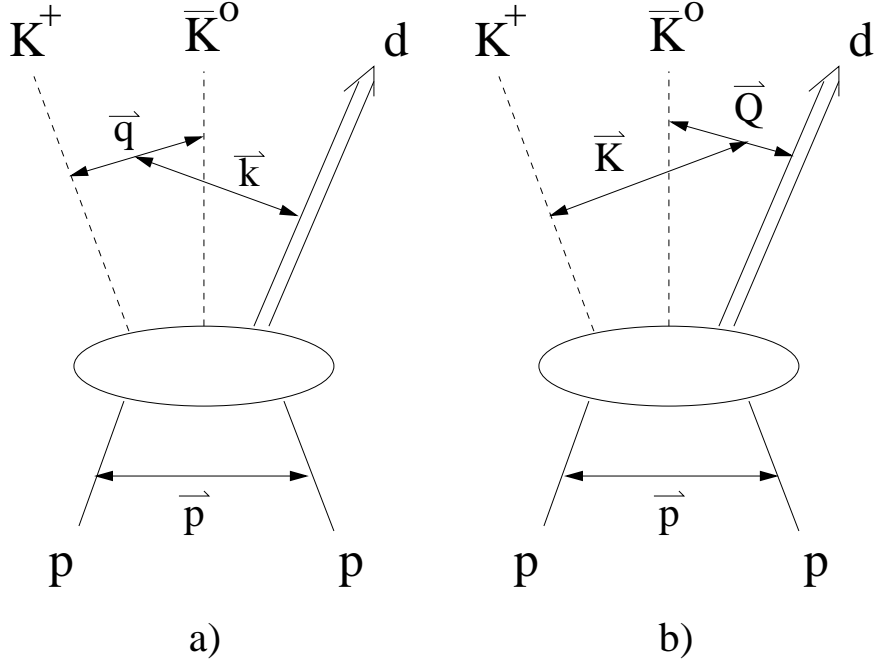}}
  \caption{Illustration of the coordinate system used in the analysis
for the reaction $pp\to d\bar K^0 K^+$. }
\label{koor}
\end{figure}

The data for three different angular distributions are shown in Fig.
\ref{andist}, those for two invariant mass distributions in Fig. \ref{mdist}.
Note, the error bars are dominated by statistics!
The data were fitted based on the assumption that either the $K\bar K$ system is
in an $s$--wave while the deuteron with respect to this system is in an
$p$--wave $or$ the $K\bar K$ system is in a $p$--wave while the deuteron with
respect to this system in in a $s$--wave (as shown in the previous section both
subsystems in an $s$--wave simultaneously is not allowed).  These assumptions
lead to the following ansatz for the square of the spin averaged matrix
element \cite{a0exp}\footnote{Due to the proximity of the kaon threshold we
  use non--relativistic kinematics.}:
\begin{equation}
\bar{|M|^2} = C_0^qq^2+C_0^kk^2+C_1(\vec k\cdot \hat p)^2+C_2(\vec q\cdot \hat p)^2
+C_3(\vec k\cdot \vec q)^2+C_4(\vec k\cdot \hat p)(\vec q\cdot \hat p) \ ,
\label{mform}
\end{equation}
where $\hat p = \vec p /|\vec p |$ denotes the direction of the beam, $\vec
k=\vec p_d$ is the relative momentum of the deuteron with respect to
the $K\bar K$ system that at the same time agrees to the cms momentum of the
deuteron,
and $\vec q = (\vec p_{K^+}-\vec p_{K^0})/2$ denotes the relative momentum of
the kaons. The coordinate system is illustrated in Fig. \ref{koor}a).
A fit  to the data  was carried out  
 and the
 parameters found are given in Table \ref{cpara}. The result of the
fit is shown in Figs. \ref{andist} and \ref{mdist}. The figures show
not only the result of the overall fit of all parameters (solid line)\footnote{Please note that
the distributions given in Ref. \cite{a0exp} are not sensitive to 
$C_3$ and $C_4$ individually, but to the combination $C_3+(1/3)C_4$ only.},
but also the fraction of the total result that stems from purely $K\bar K$
$s$--waves
(dashed--line) and from purely $K\bar K$
$p$--waves (dotted--line). The data clearly show, that the dominant
fraction of the data stem from   $K\bar K$
$s$--waves. 

\begin{table}[!t]
\begin{tabular}{ccccc} 
\hline
{$C_0^q$} &
{$C_0^k$} & 
{$C_1$} &
{$C_2$} &
{$C_3+\frac{1}{3}C_4$}\\
\hline
$0\pm 0.1$ &$1 \pm 0.03$ &$-0.6\pm 0.1$ &$1.26 \pm 0.08$ &$-0.36\pm 0.17$\\
\hline
\end{tabular}
\caption{Result for the $C$ parameters from a fit to the experimental data.
The parameters are given in units of $C_0^k$.}
\label{cpara}
\end{table}

To make the latter statement more quantitative one may use the contribution of
the $K\bar K$ $s$--waves to the total cross section as a measure of their
significance. This leads to  the following contributions:
$$
\mbox{$K\bar K$ $s$--waves} \ \ 83 \ \% \qquad \mbox{and} \qquad \mbox{$K\bar
  K$ $p$--waves} \ \ 17 \% \ .
$$
Since the pole that corresponds to the $a_0^+$ is located close to the
$\bar K K$ threshold, it will govern the propagation of the kaon pair in the
$s$--wave.  This is why in Ref. \cite{a0exp} it was claimed, that indeed the
production of $a_0^+$ was measured. In this context it is interesting to note
that the 17 \% of $K\bar K$ $p$--waves can be quantitatively understood as
stemming from a $\pi \pi$ meson exchange current \cite{leonid}.

Since the excess energy in the $\bar K K d$ system produced is
of the order of the width of the $a_0$, the latter does not show up as a bump
in the invariant mass distributions \cite{tarasov} and its relevance can only
be read of a stronger population of the corresponding partial wave. Note that, given
the dominance of $K\bar K$ $s$--waves, the $a_0$ should show up as a clear bump
in the invariant mass spectrum for $pp\to d\eta\pi^+$, for here the
corresponding threshold is far away.

In Ref. \cite{oset} it was argued that due to the resonant behavior of the
$\bar K N$ system---the hyperon resonance $\Lambda (1405)$ is situated close
to the $\bar K N$ threshold---the $\bar K d$ final state interaction should
play a significant role in the reaction dynamics. This final--state interaction
should lead to an enhancement of the $\bar K d$ $s$--wave. To see the amount
of $\bar K d$ $s$--waves in the total cross section we have to change to 
coordinates
that explicitly contain the $\bar K d$ relative momentum, as illustrated in
Fig. \ref{koor}b. Straight forward evaluation gives
$$
\vec k = \vec Q - \alpha \vec K \qquad \mbox{and} \qquad \vec q
=\frac{1}{2}((2-\alpha)\vec K + \vec Q) \ ,
$$
where $\alpha = m_d/(m_d+m_{\bar K})$. Given these expressions it is straight
forward to re-express Eq. (\ref{mform}) in terms of $\vec K$ and $\vec Q$
\begin{equation}
\bar{|M|^2} = B_0^QQ^2+B_0^KK^2+B_1(\vec K\cdot \hat p)^2+B_2(\vec Q\cdot \hat p)^2
+B_3(\vec K\cdot \vec Q)^2+B_4(\vec K\cdot \hat p)(\vec Q\cdot \hat p) \ ,
\label{mform2}
\end{equation}
where the various $B$--coefficients appearing are linear combinations of the 
$C$--coefficients of Eq. (\ref{mform}). E.g. 
\begin{eqnarray}
B_0^Q = \frac{1}{4} C_0^q+C_0^k+\frac{1}{2}C_3 \qquad \mbox{and} \qquad
B_2 = \frac{1}{4} C_2+C_1+\frac{1}{2}C_4 \ .
\end{eqnarray}
using the parameters of table \ref{cpara} one finds the following contributions
to the total production cross section:
$$
\mbox{$\bar K d$ $s$--waves} \ \ 54 \ \% \qquad \mbox{and} \qquad \mbox{$\bar
K  d$ $p$--waves} \ \ 46 \% \ .
$$
Thus the $\bar K d$ interaction does not play a prominent role
in the reaction $pp\to d\bar K^0 K^+$ close to the threshold.

\section{Summary and outlook}

In this manuscript and the corresponding talk we stressed that
\begin{itemize}
\item $f_0-a_0$ mixing is a very interesting phenomenon to study,
\item there is currently no unambiguous experimental signal for the existence
  of $f_0-a_0$ mixing,
\item scalar meson production in $NN$ and $dd$ collisions are especially
  suited for this research,
\item and first measurement of the reaction $pp\to d\bar K^0 K^+$ demonstrates
the prominent role of scalar mesons in these production reactions.
\end{itemize}
For details on the current experimental program at the accelerator COSY we
refer the interested reader to Ref. \cite{buscher}.

\begin{theacknowledgments}
  Special thanks to A. Kudryartsev, V. Tarasov, J. Haidenbauer, and J. Speth
  for the collaboration that led to most of the results discussed here and to
  M. B\"uscher and V. Kleber for stimulating discussions and careful reading
  of the manuscript. I also would like to thank A. H. Fariborz for a workshop
  that scientifically as well as socially was extremely enjoyable.
\end{theacknowledgments}

%
\bibliography{refs.bib}

\end{document}